\documentclass[10pt,conference]{IEEEtran}
\IEEEoverridecommandlockouts

\usepackage{cite}
\usepackage{amsmath,amssymb,amsfonts}
\usepackage{graphicx}
\usepackage{textcomp}
\usepackage{xcolor}
\usepackage{booktabs}
\usepackage{colortbl}
\usepackage{url}
\usepackage{placeins}
\usepackage[hidelinks]{hyperref}
\newcommand{\repositoryURL}{\mbox{\normalfont\ttfamily\href{https://github.com/cckkq/A11yDetectBench}{https://github.com/cckkq/A11yDetectBench}}}

\definecolor{groupShade}{HTML}{F2F4F7}
\definecolor{selectedShade}{HTML}{F6F7F9}
\definecolor{recallShade}{HTML}{EEF3F8}
\definecolor{precisionShade}{HTML}{F7F1E8}
\definecolor{interventionShade}{HTML}{F8FAF4}
\definecolor{budgetShade}{HTML}{F8FAF4}
\newcommand{\besthit}[1]{\textbf{#1}}
\newcommand{\rqsummary}[2]{%
\par\smallskip\noindent
\begingroup
\setlength{\fboxsep}{4pt}%
\colorbox{groupShade}{\parbox{\dimexpr\linewidth-2\fboxsep\relax}{\textbf{#1 summary.} #2}}%
\endgroup
\par\smallskip}

\def\BibTeX{{\rm B\kern-.05em{\sc i\kern-.025em b}\kern-.08em
    T\kern-.1667em\lower.7ex\hbox{E}\kern-.125emX}}

\title{Can Browser Agents Detect Accessibility Issues? An Interaction-Aware Benchmark for Web Accessibility Testing}

\author{
\IEEEauthorblockN{Kaiquan Chen\IEEEauthorrefmark{1}, Hang Dong, Bo Qiao, Si Qin, Qingwei Lin, Saravan Rajmohan}
\IEEEauthorblockA{\textit{Microsoft}\\
ckq0714@gmail.com, hangdong@microsoft.com, boqiao@microsoft.com\\
si.qin@microsoft.com, qlin@microsoft.com, saravar@microsoft.com}
\thanks{\IEEEauthorrefmark{1}This work was completed during the author's internship at Microsoft.}
}

\begin{document}
\maketitle
\begin{abstract}
Accessibility testing is an essential part of web application quality assurance. Automated checkers cover many programmatically detectable violations, but many assisted and manual checks still require rendered-state inspection, interaction, target identification, and judgment against accessibility requirements. Browser agents can observe and operate realistic web applications, making them a plausible substrate for interaction-aware accessibility testing. Existing browser-agent evaluations, however, mainly measure general task completion, leaving open whether agents can produce accessibility issue reports that match human review, avoid over-reporting, and cover the requested scope. We introduce A11yDetectBench, a benchmark for evaluating browser agents on accessibility issue reporting. Built on WebArena's reproducible web-application environments, A11yDetectBench draws its Assessment categories and check guidelines from Accessibility Insights for Web, an open-source tool whose Assessment workflow organizes WCAG-based manual checks. It converts human-reviewed accessibility records into executable tasks that fix a page state, target scope, and check scope, and asks the agent to return a structured issue report or no-issue verdict. The benchmark contains 686 executable task definitions across Shopping, Shopping Admin, GitLab, and Reddit, covering selected reviewed issues, no-issue controls generated from reviewer-checked repairs, and issue-list reporting. Using browser-use as a representative open-source browser-agent framework, we find that current agents are not yet reliable accessibility issue reporters. They can report some reviewed issues when the target and check guidelines are explicit, but performance drops when the relevant targets are not given in the prompt, the check scope broadens, or the task requires an issue list without extra unmatched reports. Diagnostic analyses suggest that failures are not explained only by missing accessibility-rule knowledge; agents also struggle to acquire page and interaction evidence and connect it to the right targets. A11yDetectBench makes these failures measurable and points to evidence acquisition and target identification as key directions for browser-agent accessibility testing.
The benchmark tasks and evaluation code are available at \repositoryURL.
\end{abstract}

\begin{figure*}[!t]
\centering
\includegraphics[width=\textwidth]{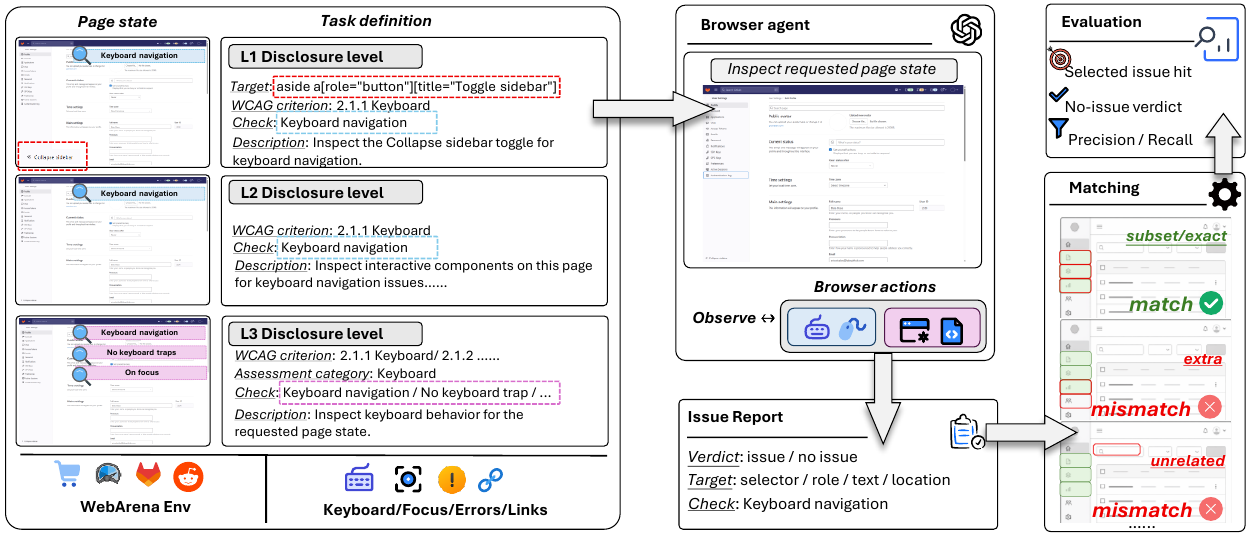}
\caption{Problem setting for browser-agent accessibility issue reporting. The task specifies a page state, a target scope, and a check scope. The agent inspects the corresponding interface state and returns a structured issue report or no-issue verdict, which the evaluator compares with the reviewed record or reviewed issue list for that task.}
\label{fig:teaser}
\end{figure*}

\section{Introduction}

Accessibility testing is a practical software-quality problem for web applications. Automated checkers can find many violations that are detectable from markup, static DOM properties, or other programmatically exposed signals. Other checks remain review-intensive because the relevant evidence depends on rendered visual state, interaction behavior, dynamic updates, focus movement, form submission, or target-specific context. A reviewer may therefore need to inspect the page as rendered, exercise the relevant interaction, identify the interface target under review, and judge that target against an accessibility requirement.

These review-intensive checks define a different kind of browser task from ordinary task completion. In this paper, we use \emph{browser agents} to mean agents that observe and act through a browser interface; they are a web-focused subset of broader GUI or computer-use agents. A typical browser-agent task asks the agent to satisfy an instruction, such as reaching a final state, answering a question, or completing an application workflow. Accessibility testing instead asks whether a rendered, and often interaction-dependent, interface state satisfies an accessibility check for a target or target set. The hard part is not only operating the browser, but also deciding what evidence is relevant, which interface target that evidence concerns, and whether the requested check is satisfied or violated; in some cases, the correct outcome is that reviewers did not find the requested issue. Browser agents bring relevant operational capabilities because they can observe screenshots, DOM or accessibility-tree serializations, browser state, and action histories, and can issue clicks, text entry, navigation, scrolling, and inspection actions in realistic interfaces. A growing body of browser-agent and web-agent benchmarks, together with broader GUI-agent benchmarks, measures these capabilities by evaluating whether an agent completes a specified task \cite{worldofbits,webshop,mind2web,webarena,visualwebarena,webvoyager,webagent,seeact,weblinx,osworld,androidwild}. This progress makes browser agents a plausible substrate for interaction-aware accessibility testing, but it does not show whether they have the accessibility-specific knowledge, interaction evidence, and target-level judgment needed for accessibility issue reporting.
 
We introduce A11yDetectBench to make this question measurable. The benchmark is built within established web accessibility standards and tool-supported review practices. The Web Content Accessibility Guidelines (WCAG) provide the normative basis for perceivable, operable, understandable, and robust web content, and the Website Accessibility Conformance Evaluation Methodology (WCAG-EM) describes evaluation methodology \cite{wcag22,wcagem}. Accessibility Insights for Web is an open-source accessibility testing tool whose Assessment feature organizes WCAG-based manual checks into categories and check guidance, including checks such as Link Function, Keyboard navigation, Error Identification, and Visible focus \cite{accessibilityinsightsweb}. A11yDetectBench uses this Assessment workflow and guidance to choose and phrase tasks, while reviewed issue and no-issue records come from assisted human review.

A11yDetectBench standardizes browser-agent accessibility issue reporting by specifying the page state, target scope, and check scope for each task. A task starts from a representative WebArena page state. Its target scope may be a named element, a target group, or a candidate set of controls; its check scope may be a specific Assessment check or a broader Assessment category. The agent returns a structured issue report or no-issue verdict, and the evaluator compares the final report with reviewed records. Figure~\ref{fig:teaser} summarizes how these task fields constrain what the agent inspects and what the evaluator checks. This formulation deliberately avoids whole-site WCAG conformance evaluation. Instead, it defines a reviewable issue-reporting task with three complementary task families: selected-issue reporting, repaired no-issue controls checked by reviewers, and issue-list reporting.

We instantiate A11yDetectBench on four application-like WebArena environments: Shopping, Shopping Admin, GitLab, and Reddit. We focus on four WCAG-based Assessment categories---Links, Keyboard, Errors/status, and Focus---because they require a mixture of semantic, DOM/AX, visual, and interaction evidence while remaining reviewable in selected WebArena page states and task scopes. The benchmark contains 686 executable task definitions: 363 selected-issue task definitions generated from 121 manually reviewed accessibility issues across three disclosure levels, 121 no-issue control definitions generated from reviewer-checked repairs, and 202 issue-list task definitions. The disclosure levels separate difficulty along target and check information: L1 discloses both the target and the specific check, L2 keeps the check fixed but hides the exact target, and L3 gives only a broader Assessment category, so the agent must identify both the concrete check and the relevant target.

Using browser-use, a representative open-source browser-agent framework, our evaluation shows that current agents are not yet reliable accessibility issue reporters. They can report some reviewed issues when the relevant target and check are explicit, but selected-issue success does not translate into complete and precise issue lists when targets are hidden or the check scope is broader. The no-issue controls show that the benchmark also measures incorrect issue reports on reviewer-checked controls. Diagnostic analyses suggest that task-relevant page and interaction observations help on Focus and Keyboard subsets, whereas adding browser steps or standard-reference text does not provide a consistent improvement. These results point to task-relevant evidence acquisition and target identification as central bottlenecks, not only knowledge of accessibility rules.

This paper makes the following contributions:
\begin{itemize}
    \item We formulate browser-agent accessibility issue reporting as a report-level testing problem. Each task specifies a page state, target scope, and check scope, and the agent returns either a structured issue report or a no-issue verdict. The evaluation compares these outputs with reviewed records through selected-issue, no-issue control, and issue-list tasks.
    \item We construct A11yDetectBench, a reproducible benchmark suite on four application-like WebArena environments and four WCAG-based Assessment categories. A11yDetectBench contains 686 executable task definitions across multiple disclosure levels, including selected-issue tasks generated from 121 manually reviewed issues, no-issue controls generated from reviewer-checked repairs, and issue-list tasks.
    \item We evaluate current browser agents with A11yDetectBench and diagnostic analyses. The results show that agents can report some reviewed issues when target and check information is explicit, but this does not translate into complete and precise issue lists under broader task scopes. The diagnostic analyses point to task-relevant evidence acquisition and target identification as central bottlenecks, highlighting concrete directions for improving browser agents for accessibility testing.
\end{itemize}

\section{Background}
\label{sec:problem-formulation}

Web accessibility testing is anchored in established standards and review workflows. WCAG defines normative requirements for accessible web content, and WCAG-EM describes how evaluations are scoped and reported \cite{wcag22,wcagem}. In practice, testing tools combine automated checks with guided manual review. Accessibility Insights for Web is an open-source accessibility testing tool whose Assessment workflow organizes WCAG-based manual checks into Assessment categories and check guidance \cite{accessibilityinsightsweb}. We use that workflow and guidance to select and phrase benchmark checks, while reviewed issue and no-issue records are produced by assisted human review.

For this benchmark, an accessibility issue-reporting task is defined by three scope fields: page state, target scope, and check scope. The page state specifies the rendered or interaction-reached state to inspect; the target scope identifies the element, target group, or candidate set under review; and the check scope specifies either a specific check or a broader Assessment category. A valid report must answer that scope by identifying the relevant target, applying the intended check, and providing observations that make the issue or no-issue verdict reviewable. This framing studies accessibility issue reporting for specified page states and task scopes.

\begin{figure*}[!t]
\centering
\includegraphics[width=\textwidth]{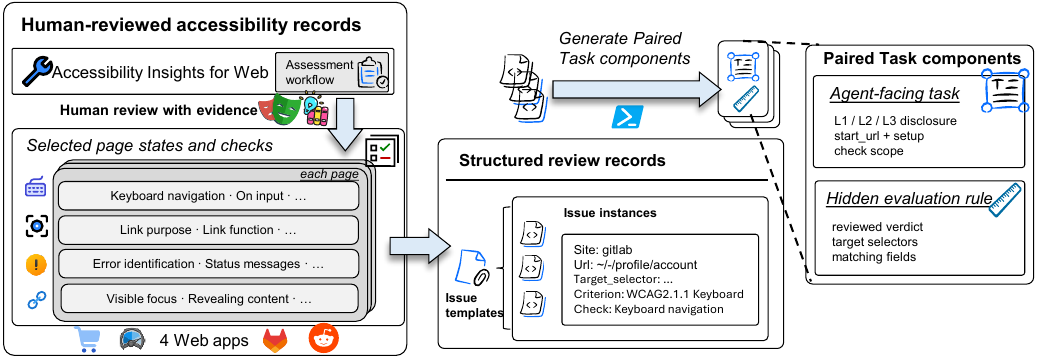}
\caption{Benchmark construction pipeline from human-reviewed accessibility records to agent-facing task definitions and hidden evaluation rules. The task definition controls disclosed page, target, and check information, while the hidden evaluation rule stores reviewed verdicts, target descriptors, and matching fields. }
\label{fig:pipeline-evaluator}
\end{figure*}

\section{Benchmark Design}
\label{sec:benchmark-design}

\subsection{Overview}

A11yDetectBench evaluates whether a browser agent can produce an accessibility issue report for a specified review scope. Each task fixes three parts of that scope: a page state, a target scope, and a check scope. The page state is the rendered or interaction-reached state to inspect. The target scope may name a single element, a group of related elements, or a candidate set of controls. The check scope may name a specific Assessment check or a broader Assessment category. Given this scope, the agent returns either a structured issue report or a no-issue verdict.

Figure~\ref{fig:pipeline-evaluator} summarizes the benchmark construction pipeline. We first select checks from the Assessment workflow. We then review WebArena page states for those checks and record the page state, target or target group, check, observed behavior, reviewed verdict, and review rationale. These reviewed records are used to generate task definitions and evaluation rules. A task definition contains only the information shown to the agent, while the evaluation rule contains the reviewed information used for matching the final report. Keeping these two components separate lets A11yDetectBench vary how much target and check information the prompt reveals without exposing reviewed verdicts, selectors, or review rationales.

The Assessment workflow contributes the categories and check guidance used to phrase tasks, while reviewed issue and no-issue decisions come from human review. WCAG supplies the normative standard, and World Wide Web Consortium (W3C) Web Accessibility Initiative (WAI) Understanding documents, review records, and site-specific rationales are used during curation and adjudication. They are not inserted as case-specific hints in the main prompt setting, so the benchmark measures what agents can report from the disclosed task scope and their browser observations.

\subsection{Check and Site Selection}

We select checks for repeatable browser-agent evaluation rather than for complete coverage of the Assessment workflow. A check is included only if it can be turned into a WebArena task with a stable review scope. This requires three properties. First, the relevant page state must be reachable from a WebArena start point, including any required login, setup step, or interaction. Second, the relevant evidence must be observable through the agent's browser interface, such as the DOM, accessibility tree, rendered page, text content, navigation behavior, or interaction behavior. Third, the reviewed decision must be verifiable from the recorded page state, target, check, observed behavior, and review rationale. We exclude checks that require whole-site conformance judgments, broad consistency checks across many pages, media-content review, user studies, or unstable runtime states.

Under these criteria, A11yDetectBench covers four WCAG-based Assessment categories that represent common manual-review areas in web accessibility testing. Links covers Link Function and Link Purpose, where reviewers judge whether link text, accessible names, destinations, activation behavior, and surrounding context make the link's purpose understandable. Keyboard covers Keyboard navigation, On Input behavior, and Character Key Shortcuts, which are central operability checks for users who navigate without a pointing device. Errors/status covers Error Identification, Error Suggestion, and Status Messages, where the key question is whether validation errors and dynamic status updates are communicated in the rendered interface and programmatically exposed state. Focus covers Visible focus and Revealing content cases, which require inspecting focus movement, focus indicators, and interaction-triggered content. Together, these categories require semantic, DOM/AX, visual, and interaction observations while still allowing repeatable judgments in selected WebArena page states.

We instantiate these checks on four WebArena applications: Shopping, Shopping Admin, GitLab, and Reddit. The environments cover consumer e-commerce, back-office administration, developer collaboration, and forum discussion. We do not treat them as representative of the full web. We use them because they are realistic, stateful, self-hosted applications in which an agent must inspect authenticated pages, repeated UI patterns, dynamic menus, submitted forms, and interaction-dependent states. Within each site, reviewers worked from site-specific page lists rather than arbitrary URL sampling. These lists included main entry pages; common list, detail, form, account, and settings pages; and a small number of deeper states needed to expose authenticated or interaction-dependent controls. Executable tasks then use reviewed page states that were reachable, repeatable, observable through the browser interface, and relevant to at least one selected Assessment check. Because the same application route can expose different interface states after setup or interaction, each benchmark task specifies the page state to inspect rather than only a URL.

Table~\ref{tab:benchmark-counts} reports executable task definitions by Assessment category, task type, and disclosure level. The unit is one task definition, not one reviewed issue or one unique page state. The selected-issue columns match across L1--L3 because each of the 121 reviewed issues yields three prompt variants with the same hidden evaluation rule but different target/check disclosure. The no-issue column contains one target-and-check disclosed repaired control per reviewed issue. The issue-list columns count page-state-and-check-scope tasks, so their L2 and L3 totals are lower and need not match the selected-issue counts.

\begin{table}[!t]
\caption{Executable task definitions by Assessment category, task type, and disclosure level.}
\label{tab:benchmark-counts}
\centering
\footnotesize
\setlength{\tabcolsep}{3.4pt}
\renewcommand{\arraystretch}{1.16}
\begin{tabular}{@{}lrrrrrrr@{}}
\toprule
\textbf{Assessment category} & \multicolumn{3}{c}{\textbf{Selected issue}} & \multicolumn{1}{c}{\textbf{No-issue}} & \multicolumn{2}{c}{\textbf{Issue-list}} & \textbf{Total} \\
\cmidrule(lr){2-4}\cmidrule(lr){5-5}\cmidrule(lr){6-7}\cmidrule(l){8-8}
 & \textbf{L1} & \textbf{L2} & \textbf{L3} & \textbf{Controls} & \textbf{L2} & \textbf{L3} & \\
\midrule
Links & 51 & 51 & 51 & 51 & 42 & 36 & 282 \\
Keyboard & 28 & 28 & 28 & 28 & 27 & 23 & 162 \\
Errors/status & 26 & 26 & 26 & 26 & 25 & 21 & 150 \\
Focus & 16 & 16 & 16 & 16 & 14 & 14 & 92 \\
\midrule
\rowcolor{groupShade}
\textbf{Total} & \textbf{121} & \textbf{121} & \textbf{121} & \textbf{121} & \textbf{108} & \textbf{94} & \textbf{686} \\
\bottomrule
\end{tabular}
\end{table}

\subsection{Human Review Process and Records}

Human review produces the records from which benchmark tasks are generated. Three reviewers conducted the review, each with more than two years of accessibility testing experience. Every candidate case was independently checked by two reviewers. A case was included only after both reviewers agreed on the page state, target or target group, check, verdict, and rationale; disagreements were discussed until the reviewers reached a shared decision.

Reviewers used the Assessment workflow as the organizing checklist and performed the actual judgment in the browser. They could operate the page freely, trigger interactions, inspect elements, and refer to DOM and accessibility-tree information when it helped determine labels, roles, target identity, or interaction behavior. For each included issue, the retained record stores the site, page state, target or target group, check, observed behavior, reviewed verdict, and observations used to justify the decision. The retained evidence includes target descriptors such as selectors, accessible names, visible text, roles, or nearby context, together with review notes needed to make the decision inspectable.

For evaluation, each reviewed issue is represented by an evaluation rule that records the reviewed verdict, target descriptors, check information, observed behavior, issue type, and matching instructions. These rules are derived from the retained review records and are hidden from the agent during task execution.

Some reviewed issues occur as repeated UI patterns. For example, the same link-purpose failure may appear across several grid action links, or the same keyboard-reachability failure may appear across several similar controls. We therefore distinguish an issue pattern from its concrete occurrences. The pattern records the repeated accessibility problem, while each occurrence records the specific target element or target group needed for matching. This prevents a repeated UI pattern from dominating selected-issue sampling while preserving the concrete targets needed for selected-issue and issue-list evaluation.

No-issue controls are constructed with the same target discipline. For an issue-present case, reviewers manually repaired the target or target set and then checked the repaired version to ensure that the specified issue was no longer present. A no-issue control therefore identifies a repaired target or target set for which reviewers confirmed that the specified issue was absent under the same check. These controls are not arbitrary passing examples and do not certify that a page is fully accessible. They test whether an agent can return a no-issue verdict when the prompt resembles an issue-present task but the specified issue has been removed for that target or target set. In this benchmark, no-issue controls are evaluated only when both the target and check are disclosed; we do not include L2- or L3-style no-issue tasks that hide the target or broaden the check scope.

\subsection{Task Definitions }

Each executable task pairs a task definition shown to the agent with an evaluation rule hidden from the agent. The task definition contains the WebArena start point, any setup instruction needed to reach the inspected state, the check or Assessment category disclosed in the prompt, the target description when the disclosure level includes it, and the required output schema. The evaluation rule contains the reviewed information used to score the final report.
A11yDetectBench separates task type from disclosure level. The task type defines what the agent is asked to return and how the output is evaluated. The disclosure level defines how much target and check information the prompt reveals. This separation lets the benchmark ask the same underlying accessibility question under easier or harder prompt conditions without changing the reviewed record used for evaluation. In practice, the same reviewed record can be tested under prompts that make target and check information progressively less explicit.

The benchmark contains three task types. Selected-issue reporting tasks ask whether the agent reports one selected reviewed issue. The page may contain other issues, but the task is scored against that selected issue. No-issue controls ask whether the agent returns a no-issue verdict for a reviewer-checked repaired target or target set where the specified issue has been removed. Issue-list reporting tasks ask the agent to list all reviewed issues for a requested page state and check scope; their evaluation rules can contain multiple reviewed issues, including same-page issues that were not sampled as selected-issue tasks.

The L1--L3 disclosure levels control target and check visibility. L1 discloses both the target or target group and the specific check, so the agent checks a specified target under a specified accessibility requirement. L2 discloses the specific check but not the exact target, so the agent must identify the relevant elements or interactions in the page state for that check. L3 discloses only a broader Assessment category, such as Links or Focus, so the agent must identify both the concrete check and the relevant target. Selected-issue reporting tasks are generated at all three levels: for each reviewed issue, A11yDetectBench creates one L1, one L2, and one L3 prompt variant, all scored against the same selected reviewed issue. No-issue controls are evaluated only in the target-and-check disclosed setting because they are designed to measure over-reporting on a repaired target, not open-ended no-issue discovery. Issue-list reporting tasks are evaluated at L2 and L3 because their purpose is to test whether the agent can recover the relevant target set, and at L3 also the concrete check, for the requested page state and scope.

As a concrete example, a reviewed Keyboard issue for the GitLab \texttt{Collapse sidebar} control yields three selected-issue task variants. L1 names the \texttt{Collapse sidebar} control and the Keyboard navigation check. L2 keeps the Keyboard navigation check but asks about keyboard reachability on the page without naming the exact target. L3 gives only the broader Keyboard Assessment category. Across these variants, the evaluation rule determines which target, behavior, and issue type count as a match. Separately, issue-list tasks are constructed from reviewed issue lists for a page state and check scope. An L2 issue-list task for that page state asks for all reviewed issues under a specific check, while an L3 issue-list task asks for all reviewed issues under the broader Assessment category. These issue-list tasks are counted by page-state-and-scope combinations, not by the number of individual reviewed issues they contain.

\subsection{Report Matching}
The evaluator scores the agent's final structured report, not the action trace. The trace is preserved for audit and failure analysis, but a run receives credit only when the final report contains a matchable issue report or the required no-issue verdict. If the trace observes an issue but the final answer does not report it, the run is not counted as a hit. Similarly, a final answer that names a plausible issue is not credited unless its structured fields match the evaluation rule.

A reported issue \(\hat{i}\) matches a reviewed issue \(i\) only when three conditions hold: the reported target matches the reviewed target, the reported check matches the requested check scope, and the reported issue type and observed behavior are compatible with the reviewed behavior or issue type. We write this as
\[
\mathrm{match}(\hat{i},i)=M_{\mathrm{target}}(\hat{i},i)\land M_{\mathrm{check}}(\hat{i},i)\land M_{\mathrm{issue}}(\hat{i},i),
\]
where \(M_{\mathrm{target}}\) checks target descriptors such as selectors, accessible names, visible text, roles, or nearby context; \(M_{\mathrm{check}}\) checks the reported check against the requested check scope; and \(M_{\mathrm{issue}}\) checks whether the reported issue type and observed behavior are compatible with the reviewed issue. The evaluator parses and preserves evidence or observation fields when present, but free-form evidence text is not used as an independent semantic judgment signal. As a result, the automatic score depends on structured target, check, and issue fields rather than open-ended interpretation of explanatory text.

Each task type uses this matching result differently. For selected-issue reporting tasks, the run is credited when at least one reported issue matches the selected reviewed issue. For no-issue controls, the correct response is a no-issue verdict for the reviewer-checked repaired target or target set where the specified issue has been removed; a reported issue for that target and check is treated as an incorrect issue report. For issue-list tasks, the evaluator matches reported issues to the reviewed issue list for the requested page state and check scope. Each reported issue can match at most one reviewed issue. Matched issues contribute to recall, while reported issues that do not match any reviewed issue in the requested task scope are false-positive reported units for precision.

\section{Evaluation Protocol and Experimental Setup}
\label{sec:evaluation-protocol}

To assess what current browser agents can do on A11yDetectBench, we run browser-use as a representative open-source browser-agent framework and compare models and diagnostic variants through the same task and scoring pipeline. This section explains how these runs are configured, how final structured reports are scored, how repeated runs are aggregated, and how the empirical questions map to result tables. Browser traces are retained for audit and failure analysis, but the reported metrics are computed from the final structured reports.

\subsection{Task Execution and Agent Setting}
\label{sec:experiments}
We instantiate each run with browser-use, which provides the observation/action loop for the evaluated agents \cite{browseruse}. A browser agent starts from a specified WebArena page state, observes the page through the browser interface, and acts through browser operations such as clicking, typing, navigation, inspection, and scrolling. Scrolling and intermediate page exploration are allowed, but they are not evaluated directly; the reported metrics use only the final structured report.

For each task, the runner restores the required WebArena state, including any login, setup action, or interaction needed to reach the reviewed page state. The agent receives the generated prompt and a structured answer schema with fields such as verdict, target, check, reported issue, and observations. The prompt includes only the information allowed by the task's disclosure level: the page state or start point, target information when disclosed, the specific Assessment check or Assessment category, the associated compact WCAG criterion statement when applicable, and the output schema. It does not include the evaluation rule, review rationale, expected verdict, expected selectors, human review notes, or the body of the WAI Understanding page.

We use OpenAI GPT-5.2, OpenAI GPT-5.5, and Moonshot AI Kimi-K2.6 as the main model conditions. The two OpenAI models represent nearby releases from a widely used frontier model family, while Kimi-K2.6 represents a different vendor and model family commonly positioned around coding, long-context use, and agentic execution. This gives representative coverage without turning the study into a broad model leaderboard. All three models run through the same browser-use pipeline and structured-output schema, so the tables compare model behavior under a shared agent runner rather than different browser tooling.

Unless a result table states otherwise, browser-use runs use temperature 0, a 25-action step budget, headless execution, the structured answer schema, and a verified login state before each task. The main evaluation includes 363 selected-issue task definitions across L1--L3 disclosure levels, 121 no-issue control definitions, and 202 issue-list task definitions. We run each task variant three times under the same agent condition to reduce dependence on a single browser trajectory; these repeats are aggregated descriptively rather than treated as a powered estimate of run-to-run variance. Malformed outputs and execution failures are logged and counted as attempted runs unless a table explicitly states a narrower denominator.

Although the reported experiments use browser-use, the task interface is not specific to it. Another agent framework can be evaluated if it can start from the specified WebArena page state or start point, operate through its own browser observation and action interface, and return the required structured report. New pages still require human-reviewed accessibility records and evaluation rules, and non-browser-use agents require runner adapters.

\subsection{Report Scoring}
In the reported experiments, a run is one execution of one task variant under one agent condition. A task variant is a concrete prompt version of a task, determined by its task type and disclosure level. The evaluator parses the final structured report and matches its verdict, target, check, issue type, and behavior fields against the evaluation rule before assigning the score for the corresponding task type. The action trace is retained for audit and failure analysis, but only issues present in the final report are scored. If the final answer describes an issue but its structured target or check does not match the evaluation rule, the run is not credited.

For selected-issue tasks, a hit requires the reported issue to match the reviewed target, check, and observed behavior or issue type. Reporting a different real issue on the same page is not a hit for that task because the task asks whether the selected reviewed issue was reported. Table~\ref{tab:main-reporting-results} reports this hit rate across L1--L3 disclosure levels; the issue-list columns in the same table use recall and precision because they score a list of reviewed issues for the requested scope. For no-issue controls, the reviewed verdict is \texttt{no\_issue\_found} for the specified target or target set. These controls come from reviewer-checked repaired cases where the requested issue was removed for the target, target set, and check under test. A correct response returns the no-issue verdict for that case. A response that reports an issue for that target or target set is an incorrect issue report on a no-issue control. No-issue controls should not be read as full-page accessibility certification or as proof that no other accessibility issue exists.

For issue-list L2/L3 tasks, the evaluator compares the reported issue list against the reviewed issue list for the requested page state and check scope. Each reported issue can match at most one reviewed issue. After this one-to-one matching, precision and recall are computed over the expanded element-level target units described below. Duplicate matches, unmatched reported issues, wrong-target reports, wrong-check reports, and unresolved targets are recorded separately. Precision is penalized by any false-positive reported unit for the requested task scope.

The output schema asks the agent to provide observations or evidence so that the report can be reviewed, but free-form evidence text is not part of the automatic matching predicate. The automatic evaluator uses structured verdict, target, check, issue type, and issue-list fields for matching. Evidence text and action traces are preserved for audit, diagnostic analysis, and manual inspection of failures. Browser-use self-reported success labels are treated as agent output, not as reviewed verdicts.

\subsection{Metrics and Repeat Aggregation}

Metrics follow the three task types defined above, and all reported rates are computed over attempted runs unless a table states otherwise. For selected-issue tasks, we report hit rate:
\[
\mathrm{HitRate}=\frac{H}{N^{\mathrm{sel}}},
\]
where \(H\) is the number of attempted runs whose final report matches the selected reviewed issue, and \(N^{\mathrm{sel}}\) is the number of attempted selected-issue runs for that condition. All attempted runs remain in the denominator. For no-issue controls, we report no-issue verdict accuracy:
\[
\mathrm{NoIssueVerdict}=\frac{CR}{N^{\mathrm{no}}},
\]
where \(CR\) is the number of no-issue control runs in which the agent returns the reviewed no-issue verdict for the reviewer-checked target or target set, and \(N^{\mathrm{no}}\) is the number of attempted no-issue runs. The result tables use the short label No-issue verdict for this higher-is-better rate. We also track \(IR/N^{\mathrm{no}}\) as the issue-report error rate, where \(IR\) is the number of no-issue runs in which the agent reports an issue.

For issue-list tasks, we compute precision and recall over element-level units. Before computing the metrics, the evaluator expands each reviewed issue to its distinct concrete target elements and scores reported targets against those elements. Repeated reports of the same concrete element do not create additional denominator units; duplicate reports are recorded separately as diagnostics. We then compute:
\[
P=\frac{\sum_s TP_s}{\sum_s (TP_s+FP_s)},\qquad
R=\frac{\sum_s TP_s}{\sum_s (TP_s+FN_s)}.
\]
Here \(s\) is the requested task scope for an issue-list task, \(TP\) counts reviewed elements matched by reported elements, \(FP\) counts distinct reported target elements plus reported issue units whose target cannot be matched to a reviewed element for that task, and \(FN\) counts reviewed elements that were missed. For each repeat, we first pool the relevant task-scope counts and compute one precision and recall value; result tables then report the arithmetic mean of these three per-repeat values.

This element-level policy means recall measures how many individual affected elements are reported and precision penalizes extra or wrong individual targets. Malformed, missing, or evaluator-invalid attempted issue-list runs remain in the denominator: their reviewed elements count toward recall with zero matches, and their precision denominator includes one false-positive reported unit. This avoids inflating precision for runs that fail to return a usable issue list.
We preserve per-repeat outcomes for audit, and the main result tables report the mean over the three repeats. Malformed outputs and execution or evaluator failures are also reported as execution and parsing failure counts, but attempted issue-list runs are not removed from the precision and recall calculations.

\subsection{Empirical Questions and Conditions}

We use this experimental setup and scoring procedure to answer four empirical questions about current browser agents:

\begin{itemize}
    \item \textbf{RQ1. Reviewed issue reporting:} How accurately do browser agents report reviewed accessibility issues when the target and check are disclosed?
    \item \textbf{RQ2. No-issue cases:} How often do browser agents return no-issue verdicts for reviewer-checked repaired targets or target sets where the requested issue has been removed?
    \item \textbf{RQ3. Issue-list quality:} How complete and precise are browser agents' issue lists for the requested page state and check scope?
    \item \textbf{RQ4. Diagnostic analyses:} Which changes to prompt reference, enriched page information, browser access, action budget, or model affect issue-reporting performance?
\end{itemize}

RQ1 is evaluated with selected-issue tasks under target-and-check disclosure. RQ2 uses no-issue controls, which are matched repaired targets or target sets for which reviewers confirmed that the requested issue was no longer present. RQ3 uses L2/L3 issue-list tasks: L2 fixes a specific check while hiding the target set, and L3 broadens the check scope to an Assessment category. RQ4 uses diagnostic variants that keep task definitions and evaluation rules fixed where possible while changing one agent-facing factor at a time. Table~\ref{tab:main-reporting-results} reports RQ1 and the main RQ3 element-level issue-list metrics, Table~\ref{tab:no-issue-verdict} reports RQ2, and RQ4 is reported through diagnostic-analysis tables that name the changed variant. Model names and diagnostic variants are reported with the corresponding result tables. A model swap, larger action budget, enriched page-information variant, WAI-reference variant, or browser-access variant is treated as a separate agent condition. The browser-access diagnostic is reported separately in the appendix because it changes the agent interface rather than only changing prompt content or trajectory length within browser-use.

\subsection{Diagnostic Analyses}

Diagnostic analyses test potential causes of failure in the main browser-use condition; they do not replace the benchmark setting. We vary one factor at a time where possible while keeping the task, output schema, and evaluator fixed. A separate appendix diagnostic uses the same Keyboard task definitions and evaluation rules under programmatic browser access; it is not a matched browser-use baseline because it changes the agent interface.
The reference-material analysis tests whether failures are reduced by giving the agent more explicit accessibility-standard guidance. We call this the WAI-reference variant: it appends the body of the public WAI Understanding page for the relevant WCAG criterion after the compact criterion statement in the main prompt. The variant appends only the main page body, not linked secondary pages, and provides explanatory standard-reference material such as criterion intent, examples, and related guidance. It does not include case-specific review rationale.

The enriched page-information variant tests whether failures are tied to evidence acquisition. It provides page and interaction observations derived from browser-accessible state, rather than reviewed issue lists or expected answers. For Keyboard reachability, the probe compares elements that appear operable on the rendered page with elements actually reached through Tab navigation and limited arrow-key traversal for composite controls. For Visible focus, the probe follows Tab navigation and records the focused element or active descendant, rendered before/after observations, and focus-related styles to flag stops where no clear page-rendered focus indicator is observed. The agent sees only probe flags and brief factual notes, not expected verdicts, expected selectors, pass/fail labels, reviewed issue lists, or human rationales. Gains in this variant are therefore interpreted as effects of making page and interaction observations easier to acquire and link to page targets, not as a replacement for the reviewed records or evaluator.

Robustness variants vary model and action budget under the same browser-use runner when possible. These comparisons use the same metric definitions and counting rules as the main condition. We interpret longer trajectories by their effect on the same report metrics, not by the number of actions completed.

\section{Results and Analysis}
\label{sec:results}

This section answers the empirical questions in Section~\ref{sec:evaluation-protocol}. Unless a table is explicitly labeled as a diagnostic analysis, results use the standard browser-use condition with three synchronized repeats. We interpret the values descriptively rather than as model rankings, because the check categories differ in the kind of visual, semantic, and interaction evidence they require. Selected-issue and no-issue columns are run-level rates, while issue-list columns report element-level recall and precision.

\subsection{RQ1: Reviewed Issue Reporting}

Table~\ref{tab:main-reporting-results} reports the main results for selected-issue tasks and issue-list tasks under the standard browser-use condition. RQ1 focuses on the Selected-L1 Hit column, where the prompt discloses both the target and the specific check. A hit means that the final structured report matches the selected reviewed issue; reporting another issue on the same page is not credited for this metric. This is the most guided issue-reporting setting in the benchmark, because the agent receives both the target and the applicable check.

\begin{table}[!t]
\caption{Selected-issue hit rates by disclosure level and element-level issue-list recall/precision under the standard browser-use condition.}
\label{tab:main-reporting-results}
\centering
\scriptsize
\setlength{\tabcolsep}{4.0pt}
\renewcommand{\arraystretch}{1.10}
\begin{tabular}{@{}ll>{\columncolor{selectedShade}}c>{\columncolor{selectedShade}}c>{\columncolor{selectedShade}}c>{\columncolor{recallShade}}c>{\columncolor{precisionShade}}c>{\columncolor{recallShade}}c>{\columncolor{precisionShade}}c@{}}
\toprule
\textbf{Category} & \textbf{Model} & \multicolumn{3}{c}{\textbf{Selected-issue hit}} & \multicolumn{2}{c}{\textbf{Issue-list L2}} & \multicolumn{2}{c}{\textbf{Issue-list L3}} \\
\cmidrule(lr){3-5}\cmidrule(lr){6-7}\cmidrule(l){8-9}
& & \textbf{L1} & \textbf{L2} & \textbf{L3} & \textbf{Recall} & \hspace{-4pt}\textbf{Precision}\hspace{-4pt} & \textbf{Recall} & \hspace{-4pt}\textbf{Precision}\hspace{-4pt} \\
\midrule
Keyboard & GPT-5.2 & 0.82 & 0.39 & 0.14 & 0.15 & \besthit{0.55} & 0.10 & \besthit{0.51} \\
& GPT-5.5 & \besthit{0.85} & \besthit{0.46} & \besthit{0.30} & \besthit{0.44} & 0.35 & \besthit{0.38} & 0.31 \\
& Kimi-K2.6 & 0.57 & 0.36 & 0.21 & 0.42 & 0.36 & \besthit{0.38} & 0.21 \\
\addlinespace[2pt]
Focus & GPT-5.2 & 0.79 & 0.06 & 0.02 & 0.02 & 0.08 & 0.01 & \besthit{0.03} \\
& GPT-5.5 & \besthit{0.81} & \besthit{0.31} & 0.10 & 0.15 & \besthit{0.17} & 0.03 & \besthit{0.03} \\
& Kimi-K2.6 & 0.50 & 0.27 & \besthit{0.13} & \besthit{0.17} & 0.03 & \besthit{0.10} & 0.02 \\
\addlinespace[2pt]
Errors/ & GPT-5.2 & 0.60 & \besthit{0.56} & 0.36 & \besthit{0.51} & \besthit{0.61} & 0.30 & 0.34 \\
status & GPT-5.5 & \besthit{0.68} & 0.47 & 0.36 & 0.40 & 0.40 & \besthit{0.40} & \besthit{0.40} \\
& Kimi-K2.6 & \besthit{0.68} & 0.54 & \besthit{0.41} & 0.47 & 0.48 & 0.36 & 0.23 \\
\addlinespace[2pt]
Links & GPT-5.2 & 0.45 & 0.38 & 0.30 & 0.21 & 0.50 & 0.21 & 0.38 \\
& GPT-5.5 & 0.65 & 0.56 & 0.35 & 0.56 & \besthit{0.73} & 0.33 & \besthit{0.43} \\
& Kimi-K2.6 & \besthit{0.93} & \besthit{0.69} & \besthit{0.43} & \besthit{0.61} & 0.64 & \besthit{0.48} & \besthit{0.43} \\
\bottomrule
\end{tabular}

\end{table}

The selected-issue results show that current agents can report some reviewed issues when the target and check are explicit, but performance is uneven across categories. GPT-5.5 has the highest hit rate on Keyboard and Focus, Kimi-K2.6 has the highest hit rate on Links, and GPT-5.5 and Kimi-K2.6 tie on Errors/status. These category-level differences should be read with care: the checks require different evidence, and Selected-L1 credits only one reviewed issue at a time.

\rqsummary{RQ1}{When the target and check guidance are given, agents can report some reviewed issues, but this is the easiest form of the task. It does not yet show that they can carry out accessibility testing with less guidance.}

\subsection{RQ2: No-Issue Cases}

No-issue controls test whether the agent can avoid reporting an issue on repaired targets or target sets for which reviewers confirmed that the requested issue was no longer present. Table~\ref{tab:no-issue-verdict} reports no-issue verdict accuracy, the fraction of no-issue control runs on which the agent returns the reviewed no-issue verdict. Failed or missing final answers are not credited as no-issue verdicts.

\begin{table}[!t]
\caption{No-issue verdict accuracy for reviewer-checked repaired controls.}
\label{tab:no-issue-verdict}
\centering
\footnotesize
\setlength{\tabcolsep}{4.2pt}
\renewcommand{\arraystretch}{1.14}
\begin{tabular}{@{}lcccc@{}}
\toprule
\textbf{Model} & \multicolumn{4}{c}{\textbf{No-issue verdict accuracy} (0.00--1.00)} \\
\cmidrule(l){2-5}
 & \textbf{Keyboard} & \textbf{Focus} & \textbf{Errors/status} & \textbf{Links} \\
\midrule
GPT-5.2  & \besthit{0.93} & \besthit{1.00} & \besthit{1.00} & \besthit{0.94} \\
GPT-5.5  & 0.92 & 0.98 & \besthit{1.00} & 0.89 \\
Kimi-K2.6 & 0.83 & 0.69 & 0.91 & 0.92 \\
\bottomrule
\end{tabular}

\end{table}

Table~\ref{tab:no-issue-verdict} shows that no-issue behavior is not simply the inverse of selected-issue reporting. GPT-5.2 and GPT-5.5 return the reviewed no-issue verdict on most controls across all four categories: both are at or above 0.92 on Keyboard, at or above 0.98 on Focus, and both reach 1.00 on Errors/status. Links is also high for all three models, with accuracy from 0.89 to 0.94. The main drop appears for Kimi-K2.6 on the interaction-heavy Keyboard and Focus controls, especially Focus at 0.69. Those lower values mean that the agent more often reports an issue even though reviewers confirmed that the specified issue had been removed for the target or target set. The metric should therefore be read as a check on over-reporting within the reviewed scope, not as full-page accessibility certification.

\rqsummary{RQ2}{No-issue controls show that the benchmark can distinguish issue reporting from over-reporting. GPT-5.2 and GPT-5.5 are usually conservative on these controls, while Kimi-K2.6 is more likely to over-report.}

\subsection{RQ3: Issue-List Quality}

RQ3 asks whether agents can produce the output closest to a practical accessibility review: an issue list for the requested page state and check scope. The issue-list columns in Table~\ref{tab:main-reporting-results} report element-level recall and precision. L2 fixes the specific check but hides the affected target set, so the agent must find the relevant targets. L3 gives only a broader Assessment category, so the agent must also decide which concrete check applies. These settings are stricter than Selected-L1, which discloses the target and check and credits a single reviewed issue. They are also counted differently: each issue-list task may contain multiple reviewed issues, and the metric scores the resulting list rather than a single selected issue.

Table~\ref{tab:main-reporting-results} shows a clear drop from selected-issue reporting to issue-list reporting. Focus is the weakest category, with best L2 recall of 0.17, best L3 recall of 0.10, and very low precision. Links is the strongest category, but recall still drops from L2 to L3, showing that broader check scope adds difficulty even when the category is comparatively easier. Keyboard and Errors/status sit between these extremes: agents can recover some reviewed targets, but precision is mixed, so higher recall often comes with incorrect or unmatched targets. Across categories, Selected-L1 hit rates do not predict issue-list quality; agents can report a disclosed reviewed issue and still fail to assemble a complete and precise issue list when targets are hidden or the check scope broadens.

\rqsummary{RQ3}{Issue-list tasks reveal the main gap in current agents. Performance is especially weak for Focus, Links is comparatively easier, and broader L3 scopes usually reduce recall. Overall, selected-issue success does not carry over to complete and precise issue lists.}

\subsection{RQ4: Diagnostic Analyses}

RQ4 examines three diagnostic variants that may explain the failures observed in the main results: a larger action budget, probe-derived page and interaction observations, and additional WAI reference text. These analyses are narrower than the main benchmark. They keep the task, output schema, and evaluator fixed where possible, and are used to interpret failure modes rather than define new benchmark conditions.

\paragraph{Step budget}
Table~\ref{tab:step-budget-sensitivity} compares matched GPT-5.5 Keyboard and Focus runs under 25- and 50-step budgets. If the main issue were too little exploration time, the 50-step condition should improve the same metrics consistently.

\begin{table}[!t]
\caption{Step-budget diagnostic on matched GPT-5.5 Keyboard and Focus runs.}
\label{tab:step-budget-sensitivity}
\centering
\scriptsize
\setlength{\tabcolsep}{3.0pt}
\renewcommand{\arraystretch}{1.10}
\begin{tabular}{@{}lcccccc@{}}
\toprule
\textbf{Steps} & \multicolumn{1}{c}{\textbf{Selected-L1}} & \multicolumn{1}{c}{\textbf{No-issue verdict}} & \multicolumn{2}{c}{\textbf{Issue-list L2}} & \multicolumn{2}{c}{\textbf{Issue-list L3}} \\
\cmidrule(lr){2-2}\cmidrule(lr){3-3}\cmidrule(lr){4-5}\cmidrule(l){6-7}
 & \textbf{Hit} & \textbf{Accuracy} & \textbf{Recall} & \textbf{Precision} & \textbf{Recall} & \textbf{Precision} \\
\midrule
\rowcolor{groupShade}\multicolumn{7}{@{}l}{\textbf{Keyboard}}\\[-1pt]
25 & 0.85 & 0.92 & 0.44 & 0.35 & 0.38 & \besthit{0.31} \\
\rowcolor{budgetShade}
50 & \besthit{0.86} & \besthit{0.93} & \besthit{0.45} & \besthit{0.44} & \besthit{0.49} & 0.24 \\
\addlinespace[2pt]
\rowcolor{groupShade}\multicolumn{7}{@{}l}{\textbf{Focus}}\\[-1pt]
25 & \besthit{0.81} & \besthit{0.98} & \besthit{0.15} & \besthit{0.17} & 0.03 & \besthit{0.03} \\
\rowcolor{budgetShade}
50 & 0.73 & 0.96 & 0.11 & \besthit{0.17} & \besthit{0.08} & 0.02 \\
\bottomrule
\end{tabular}

\end{table}

The larger budget provides only limited and inconsistent gains. On Keyboard, 50 steps slightly improves most metrics, including L2 precision from 0.35 to 0.44 and L3 recall from 0.38 to 0.49, but L3 precision drops from 0.31 to 0.24. On Focus, the larger budget improves only L3 recall, from 0.03 to 0.08, while Selected-L1 hit and L2 recall decrease. This pattern suggests that some tasks benefit from more exploration, but failures are not mainly explained by running out of browser actions.

\paragraph{Enriched page information}
Table~\ref{tab:enriched-page-information} tests whether agents fail because the relevant page and interaction evidence is hard to obtain during ordinary browser-use runs. This paired GPT-5.2 variant adds probe-derived observations for L2 Visible focus and Keyboard navigation issue-list subsets, while still requiring the agent to decide which targets belong in the final issue list.

\begin{table}[!t]
\caption{Enriched page-information variant on paired GPT-5.2 Visible focus and Keyboard navigation L2 issue-list subsets.}
\label{tab:enriched-page-information}
\centering
\footnotesize
\setlength{\tabcolsep}{4.0pt}
\renewcommand{\arraystretch}{1.14}
\begin{tabular}{@{}lcc@{}}
\toprule
\textbf{Condition} & \multicolumn{2}{c}{\textbf{Issue-list L2}} \\
\cmidrule(l){2-3}
 & \textbf{Recall} & \textbf{Precision} \\
\midrule
\rowcolor{groupShade}\multicolumn{3}{@{}l}{\textbf{Visible focus}}\\[-1pt]
Main & 0.03 & 0.10 \\
\rowcolor{interventionShade}
Enriched page information & \besthit{0.20} & \besthit{0.18} \\
\addlinespace[2pt]
\rowcolor{groupShade}\multicolumn{3}{@{}l}{\textbf{Keyboard navigation}}\\[-1pt]
Main & 0.10 & 0.47 \\
\rowcolor{interventionShade}
Enriched page information & \besthit{0.67} & \besthit{0.88} \\
\bottomrule
\end{tabular}

\end{table}

This variant gives the clearest improvement among the diagnostics. Visible focus recall increases from 0.03 to 0.20 and precision from 0.10 to 0.18. Keyboard navigation improves more sharply, with recall moving from 0.10 to 0.67 and precision from 0.47 to 0.88. Because L2 fixes the check and hides the target set, these gains point to target finding and interaction evidence as important failure points. The scores remain imperfect, especially for Visible focus, so enriched information helps but does not solve issue-list reporting.

\paragraph{WAI reference text}
Table~\ref{tab:wai-prompt-diagnostic} asks whether adding standards-reference text helps. The WAI-reference variant appends the public WAI Understanding page for the relevant WCAG criterion on a paired GPT-5.5 WCAG 2.1.1 Keyboard subset.

\begin{table}[!t]
\caption{WAI-reference diagnostic analysis on the paired GPT-5.5 WCAG 2.1.1 Keyboard subset.}
\label{tab:wai-prompt-diagnostic}
\centering
\footnotesize
\setlength{\tabcolsep}{4.0pt}
\renewcommand{\arraystretch}{1.18}
\begin{tabular}{@{}lcccc@{}}
\toprule
\textbf{Prompt} & \multicolumn{2}{c}{\textbf{L1 tasks}} & \multicolumn{2}{c}{\textbf{Issue-list L2}} \\
\cmidrule(lr){2-3}\cmidrule(l){4-5}
 & \textbf{Selected} & \textbf{No-issue} & \textbf{Recall} & \textbf{Precision} \\
\midrule
\rowcolor{groupShade}
Main prompt & \besthit{0.94} & \besthit{0.92} & \besthit{0.43} & 0.37 \\
WAI reference & 0.73 & 0.88 & 0.38 & \besthit{0.39} \\
\bottomrule
\end{tabular}

\end{table}

The WAI-reference variant does not help on this paired subset. Selected-L1 hit decreases from 0.94 to 0.73, no-issue verdict accuracy decreases from 0.92 to 0.88, and L2 recall decreases from 0.43 to 0.38; only L2 precision changes slightly upward, from 0.37 to 0.39. This suggests that adding more rule text is not enough; agents still need to locate the relevant evidence and apply it to the right page targets.

\rqsummary{RQ4}{The diagnostics point most strongly to evidence acquisition and target selection. Extra page and interaction observations improve the paired Visible focus and Keyboard navigation L2 subsets, while more steps and more WAI reference text do not provide consistent gains.}

\section{Related Work}
\label{sec:related-work}

\paragraph{Browser, Web, and GUI-agent Benchmark}
Modern web-agent and GUI-agent benchmarks establish realistic environments for browser, mobile, and desktop operation \cite{worldofbits,webshop,mind2web,webarena,visualwebarena,webvoyager,webagent,seeact,weblinx,workarena,assistantbench,browsergym,osworld,androidwild,guiceval,pspabench}. They make agent progress measurable by checking whether an agent follows a requested instruction, produces an answer, or completes an application workflow. Recent benchmarks further refine this evaluation with safety, privacy, enterprise workflow, structured extraction, economic reasoning, software-engineering, or tool-access constraints \cite{stwebagentbench,agentdam,weblists,workarenapp,econwebarena,swebench,toolillusion}. Our work uses the same executable-web setting, but asks a software-testing question: whether a browser agent can produce an accessibility issue report that matches the reviewed record for the requested task scope.

\paragraph{Accessibility testing tools and methods}
Accessibility testing already has mature standards, tools, and review practices. WCAG, WAI-ARIA (Accessible Rich Internet Applications), accessible-name computation, ACT rules, and WCAG-EM define requirements and evaluation methodology, while Accessibility Insights for Web provides an open-source tool with automated checks, assisted checks, and a WCAG-based manual Assessment workflow \cite{wcag22,waiaria12,accname12,actrulesformat,wcagem,accessibilityinsightsweb}. Tools such as axe-core, Lighthouse, WAVE, and Pa11y provide automated coverage for many programmatically detectable issues \cite{axecore,lighthouse,wave,pa11y}. Research systems further study semantic web-accessibility issues, dynamic content, mobile and assistive-service scenarios, crawling, screen-reader transcripts, replayable tests, LLM-assisted detection, and remediation \cite{axeray,timestump,a11yscan,mate,latte,groundhog,screenaudit,taskaudit,axnav,gena11y,accessguru,automatedremediation,tabulara11y,llmweba11yslr,aaa}. These systems provide standards, checkers, evidence sources, repair pipelines, or assessment assistance. Our benchmark is different in role: it evaluates a general-purpose browser agent's reported accessibility issues against manually reviewed records.

\section{Threats to Validity}
\label{sec:threats}

The first threat concerns benchmark scope. A11yDetectBench covers four WebArena applications and a selected set of Assessment categories and checks. These applications provide reproducible interactive UI states, but they do not represent the full web, every WebArena environment, or the full WCAG and manual Assessment space. The benchmark should therefore be read as an evaluation of browser-agent issue reporting on reviewed interactive cases, not as evidence about full-page or whole-site accessibility auditing.
The second threat concerns reviewed records. The labels are tied to specified page states, targets or candidate sets, and checks. Double review and consensus discussion reduce labeling error, but reviewers may still miss issues, and unscored page regions may contain additional accessibility issues outside the requested task scope. To make these decisions inspectable, the artifact links each task definition to its evaluation rule, reviewed record, run output, and retained evidence fields where release constraints allow.
The third threat concerns the agent and scoring setup. The primary condition uses browser-use's default observation and action interface, and results may differ with other agent frameworks, browser tools, models, prompts, or action budgets. The diagnostic analyses change selected factors on narrower subsets rather than the full benchmark. The automatic evaluator scores structured report fields rather than free-form reasoning traces; traces and evidence fields are retained for audit and failure analysis, but the reported metrics measure the final issue report that the agent returns.

\section{Conclusion}
\label{sec:conclusion}

We presented A11yDetectBench, a WebArena-based benchmark for evaluating browser agents on accessibility testing. It converts human-reviewed accessibility records into executable tasks with fixed page states, target scopes, and check scopes, and separates the agent-facing task definition from the hidden evaluation rule. Our browser-use evaluation shows that current agents can report some reviewed issues when target and check information is explicit, but still struggle to avoid unsupported reports and to produce issue lists that cover the requested scope. The results point to better task-relevant evidence acquisition, target identification, and handling of no-issue cases as concrete directions for browser-agent accessibility testing.

\section*{Data Availability}
{\raggedright
The A11yDetectBench repository contains the benchmark tasks, scoring rules, and code for running and evaluating browser agents:\newline
\repositoryURL.\par}

\bibliographystyle{IEEEtran}
\bibliography{references}

\begin{thebibliography}{10}
\providecommand{\url}[1]{#1}
\csname url@samestyle\endcsname
\providecommand{\newblock}{\relax}
\providecommand{\bibinfo}[2]{#2}
\providecommand{\BIBentrySTDinterwordspacing}{\spaceskip=0pt\relax}
\providecommand{\BIBentryALTinterwordstretchfactor}{4}
\providecommand{\BIBentryALTinterwordspacing}{\spaceskip=\fontdimen2\font plus
\BIBentryALTinterwordstretchfactor\fontdimen3\font minus
  \fontdimen4\font\relax}
\providecommand{\BIBforeignlanguage}[2]{{%
\expandafter\ifx\csname l@#1\endcsname\relax
\typeout{** WARNING: IEEEtran.bst: No hyphenation pattern has been}%
\typeout{** loaded for the language `#1'. Using the pattern for}%
\typeout{** the default language instead.}%
\else
\language=\csname l@#1\endcsname
\fi
#2}}
\providecommand{\BIBdecl}{\relax}
\BIBdecl

\bibitem{worldofbits}
T.~Shi, A.~Karpathy, L.~Fan, J.~Hernandez, and P.~Liang, ``World of bits: An
  open-domain platform for web-based agents,'' in \emph{Proceedings of the 34th
  International Conference on Machine Learning}, 2017, pp. 3135--3144.

\bibitem{webshop}
S.~Yao, H.~Chen, J.~Yang, and K.~Narasimhan, ``Webshop: Towards scalable
  real-world web interaction with grounded language agents,'' in \emph{Advances
  in Neural Information Processing Systems}, 2022.

\bibitem{mind2web}
X.~Deng, Y.~Gu, B.~Zheng, S.~Chen, S.~Stevens, B.~Wang, H.~Sun, and Y.~Su,
  ``Mind2web: Towards a generalist agent for the web,'' in \emph{Advances in
  Neural Information Processing Systems}, 2023.

\bibitem{webarena}
S.~Zhou, F.~F. Xu, H.~Zhu, X.~Zhou, R.~Lo, A.~Sridhar, X.~Cheng, T.~Ou,
  Y.~Bisk, D.~Fried, U.~Alon, and G.~Neubig, ``Webarena: A realistic web
  environment for building autonomous agents,'' in \emph{International
  Conference on Learning Representations}, 2024.

\bibitem{visualwebarena}
J.~Y. Koh, R.~Lo, L.~Jang, V.~Duvvur, M.~C. Lim, P.-Y. Huang, G.~Neubig,
  S.~Zhou, R.~Salakhutdinov, and D.~Fried, ``Visualwebarena: Evaluating
  multimodal agents on realistic visually grounded web tasks,'' in
  \emph{Proceedings of the 62nd Annual Meeting of the Association for
  Computational Linguistics (Volume 1: Long Papers)}, 2024.

\bibitem{webvoyager}
H.~He, W.~Yao, K.~Ma, W.~Yu, Y.~Dai, H.~Zhang, Z.~Lan, and D.~Yu, ``Webvoyager:
  Building an end-to-end web agent with large multimodal models,'' in
  \emph{Proceedings of the 62nd Annual Meeting of the Association for
  Computational Linguistics (Volume 1: Long Papers)}, 2024.

\bibitem{webagent}
I.~Gur, H.~Furuta, A.~Huang, M.~Safdari, Y.~Matsuo, D.~Eck, and A.~Faust, ``A
  real-world webagent with planning, long context understanding, and program
  synthesis,'' in \emph{International Conference on Learning Representations},
  2024.

\bibitem{seeact}
B.~Zheng, B.~Gou, J.~Kil, H.~Sun, and Y.~Su, ``{GPT-4V(ision)} is a generalist
  web agent, if grounded,'' \emph{arXiv preprint arXiv:2401.01614}, 2024.

\bibitem{weblinx}
X.~H. L{\`u}, Z.~Kasner, and S.~Reddy, ``{WebLINX}: Real-world website
  navigation with multi-turn dialogue,'' \emph{arXiv preprint
  arXiv:2402.05930}, 2024.

\bibitem{osworld}
T.~Xie, D.~Zhang, J.~Chen, X.~Li, S.~Zhao, R.~Cao, J.~H. Toh, Z.~Cheng,
  D.~Shin, F.~Lei, Y.~Liu, Y.~Xu, S.~Zhou, S.~Savarese, C.~Xiong, V.~Zhong, and
  T.~Yu, ``{OSWorld}: Benchmarking multimodal agents for open-ended tasks in
  real computer environments,'' in \emph{Advances in Neural Information
  Processing Systems}, vol.~37, 2024, pp. 52\,040--52\,094.

\bibitem{androidwild}
C.~Rawles, A.~Li, D.~Rodriguez, O.~Riva, and T.~Lillicrap, ``Android in the
  wild: A large-scale dataset for android device control,'' in \emph{Advances
  in Neural Information Processing Systems}, vol.~36, 2023, pp.
  59\,708--59\,728.

\bibitem{wcag22}
{World Wide Web Consortium}, ``{Web Content Accessibility Guidelines (WCAG)
  2.2},'' \url{https://www.w3.org/TR/WCAG22/}, 2023.

\bibitem{wcagem}
{W3C Web Accessibility Initiative}, ``{Website Accessibility Conformance
  Evaluation Methodology (WCAG-EM) 1.0},''
  \url{https://www.w3.org/TR/WCAG-EM/}, 2014.

\bibitem{accessibilityinsightsweb}
{Microsoft}, ``{Accessibility Insights for Web},''
  \url{https://github.com/microsoft/accessibility-insights-web}, 2026, accessed
  2026-05-11.

\bibitem{browseruse}
{browser-use contributors}, ``{browser-use}: Make websites accessible for ai
  agents,'' \url{https://github.com/browser-use/browser-use}, 2024, accessed:
  2026-06-20.

\bibitem{workarena}
A.~Drouin, M.~Gasse, M.~Caccia, I.~H. Laradji, M.~Del~Verme, T.~Marty,
  L.~Boisvert, M.~Thakkar, Q.~Cappart, D.~Vazquez, N.~Chapados, and A.~Lacoste,
  ``Workarena: How capable are web agents at solving common knowledge work
  tasks?'' \emph{arXiv preprint arXiv:2403.07718}, 2024.

\bibitem{assistantbench}
O.~Yoran, S.~J. Amouyal, C.~Malaviya, B.~Bogin, O.~Press, and J.~Berant,
  ``Assistantbench: Can web agents solve realistic and time-consuming tasks?''
  in \emph{Proceedings of the 2024 Conference on Empirical Methods in Natural
  Language Processing}, 2024, pp. 8938--8968.

\bibitem{browsergym}
T.~Le~Sellier De~Chezelles, M.~Gasse, A.~Drouin, M.~Caccia, L.~Boisvert,
  M.~Thakkar, T.~Marty, R.~Assouel, S.~O. Shayegan, L.~K. Jang, X.~H. L{\`u},
  O.~Yoran, D.~Kong, F.~F. Xu, S.~Reddy, Q.~Cappart, G.~Neubig \emph{et~al.},
  ``The {BrowserGym} ecosystem for web agent research,'' \emph{arXiv preprint
  arXiv:2412.05467}, 2024.

\bibitem{guiceval}
Y.~Li, Y.~Liu, H.~Lu, Z.~Xia, H.~Wang, K.~Han, C.~Yang, J.~Wu, J.~Xu, R.~Shi,
  and Y.~Huang, ``Gui-ceval: A hierarchical and comprehensive chinese benchmark
  for mobile gui agents,'' in \emph{Proceedings of the IEEE/CVF Conference on
  Computer Vision and Pattern Recognition}, 2026, pp. 20\,303--20\,312.

\bibitem{pspabench}
H.~Nie, X.~Liu, Y.~Bai, Y.~Wang, Y.~Liu, Q.~Yao, and Z.~Wang, ``Pspa-bench: A
  personalized benchmark for smartphone gui agent,'' \emph{arXiv preprint
  arXiv:2603.29318}, 2026.

\bibitem{stwebagentbench}
I.~Levy, B.~Wiesel, S.~Marreed, A.~Oved, A.~Yaeli, N.~Mashkif, and S.~Shlomov,
  ``{ST-WebAgentBench}: A benchmark for evaluating safety and trustworthiness
  in web agents,'' in \emph{International Conference on Learning
  Representations}, 2026.

\bibitem{agentdam}
A.~Zharmagambetov, C.~Guo, I.~Evtimov, M.~Pavlova, R.~Salakhutdinov, and
  K.~Chaudhuri, ``{AgentDAM}: Privacy leakage evaluation for autonomous web
  agents,'' in \emph{Advances in Neural Information Processing Systems}, 2025.

\bibitem{weblists}
A.~Bohra, M.~Saroyan, D.~Melkozerov, V.~Karufanyan, G.~Maher, P.~Weinberger,
  A.~Harutyunyan, and G.~Campagna, ``Weblists: Extracting structured
  information from complex interactive websites using executable llm agents,''
  \emph{arXiv preprint arXiv:2504.12682}, 2025.

\bibitem{workarenapp}
L.~Boisvert, M.~Thakkar, M.~Gasse, M.~Caccia, T.~Le~Sellier De~Chezelles,
  Q.~Cappart, N.~Chapados, A.~Lacoste, and A.~Drouin, ``{WorkArena++}: Towards
  compositional planning and reasoning-based common knowledge work tasks,'' in
  \emph{Advances in Neural Information Processing Systems}, vol.~37, 2024, pp.
  5996--6051.

\bibitem{econwebarena}
Z.~Liu and Y.~Quan, ``Econwebarena: Benchmarking autonomous agents on economic
  tasks in realistic web environments,'' \emph{arXiv preprint
  arXiv:2506.08136}, 2025.

\bibitem{swebench}
C.~E. Jimenez, J.~Yang, A.~Wettig, S.~Yao, K.~Pei, O.~Press, and K.~Narasimhan,
  ``{SWE-bench}: Can language models resolve real-world github issues?'' in
  \emph{International Conference on Learning Representations}, 2024.

\bibitem{toolillusion}
R.~Lou, B.~Peng, W.~Yao, Q.~Wu, H.~Cheng, S.~Nath, W.~Yin, and J.~Gao, ``The
  tool illusion: Rethinking tool use in web agents,'' \emph{arXiv preprint
  arXiv:2604.03465}, 2026.

\bibitem{waiaria12}
{World Wide Web Consortium}, ``{Accessible Rich Internet Applications
  (WAI-ARIA) 1.2},'' \url{https://www.w3.org/TR/wai-aria-1.2/}, 2023.

\bibitem{accname12}
------, ``{Accessible Name and Description Computation 1.2},''
  \url{https://www.w3.org/TR/accname-1.2/}, 2025.

\bibitem{actrulesformat}
------, ``{Accessibility Conformance Testing (ACT) Rules Format 1.1},''
  \url{https://www.w3.org/TR/act-rules-format/}, 2022.

\bibitem{axecore}
{Deque Systems}, ``{axe-core}: Accessibility testing engine for websites and
  other html-based user interfaces,''
  \url{https://github.com/dequelabs/axe-core}, 2026, accessed 2026-06-21.

\bibitem{lighthouse}
{Google Chrome Developers}, ``{Lighthouse}: Automated auditing, performance
  metrics, and best practices for the web,''
  \url{https://developer.chrome.com/docs/lighthouse/overview}, 2026, accessed
  2026-06-21.

\bibitem{wave}
{WebAIM}, ``{WAVE}: Web accessibility evaluation tools,''
  \url{https://wave.webaim.org/}, 2026, accessed 2026-06-21.

\bibitem{pa11y}
{Pa11y Team}, ``{Pa11y}: Automated accessibility testing,''
  \url{https://pa11y.org/}, 2026, accessed 2026-06-21.

\bibitem{axeray}
M.~Bajammal and A.~Mesbah, ``Semantic web accessibility testing via
  hierarchical visual analysis,'' in \emph{2021 IEEE/ACM 43rd International
  Conference on Software Engineering (ICSE)}, 2021.

\bibitem{timestump}
F.~Mehralian, Z.~He, and S.~Malek, ``Automated accessibility analysis of
  dynamic content changes on mobile apps,'' in \emph{2025 IEEE/ACM 47th
  International Conference on Software Engineering (ICSE)}, 2025.

\bibitem{a11yscan}
Y.~Zhang, S.~Chen, X.~Xie, Z.~Liu, and L.~Fan, ``Scenario-driven and
  context-aware automated accessibility testing for android apps,'' in
  \emph{2025 IEEE/ACM 47th International Conference on Software Engineering
  (ICSE)}, 2025.

\bibitem{mate}
M.~M. Eler, J.~M. Rojas, Y.~Ge, and G.~Fraser, ``Automated accessibility
  testing of mobile apps,'' in \emph{2018 IEEE 11th International Conference on
  Software Testing, Verification and Validation}, 2018.

\bibitem{latte}
N.~Salehnamadi, A.~Alshayban, J.-W. Lin, I.~Ahmed, S.~Branham, and S.~Malek,
  ``Latte: Use-case and assistive-service driven automated accessibility
  testing framework for android,'' in \emph{Proceedings of the 2021 CHI
  Conference on Human Factors in Computing Systems}, 2021.

\bibitem{groundhog}
N.~Salehnamadi, F.~Mehralian, and S.~Malek, ``Groundhog: An automated
  accessibility crawler for mobile apps,'' in \emph{Proceedings of the 37th
  IEEE/ACM International Conference on Automated Software Engineering}, 2022.

\bibitem{screenaudit}
M.~Zhong, R.~Chen, X.~Chen, J.~Fogarty, and J.~O. Wobbrock, ``Screenaudit:
  Detecting screen reader accessibility errors in mobile apps using large
  language models,'' in \emph{Proceedings of the 2025 CHI Conference on Human
  Factors in Computing Systems}, 2025.

\bibitem{taskaudit}
M.~Zhong, X.~Chen, D.~W. Kyi, L.~Chen, J.~Fogarty, and J.~O. Wobbrock,
  ``Taskaudit: Detecting functiona11ity errors in mobile apps via agentic task
  execution,'' in \emph{Proceedings of the 2026 CHI Conference on Human Factors
  in Computing Systems}, 2026.

\bibitem{axnav}
M.~Taeb, A.~Swearngin, E.~Schoop, R.~Cheng, Y.~Jiang, and J.~Nichols,
  ``{AXNav}: Replaying accessibility tests from natural language,'' in
  \emph{Proceedings of the 2024 CHI Conference on Human Factors in Computing
  Systems}, 2024.

\bibitem{gena11y}
Z.~He, S.~F. Huq, and S.~Malek, ``Enhancing web accessibility: Automated
  detection of issues with generative ai,'' \emph{Proceedings of the ACM on
  Software Engineering}, vol.~2, no. FSE, 2025.

\bibitem{accessguru}
N.~Fathallah, D.~Hernandez, and S.~Staab, ``Accessguru: Leveraging llms to
  detect and correct web accessibility violations in html code,'' in
  \emph{Proceedings of the 27th International ACM SIGACCESS Conference on
  Computers and Accessibility}, 2025.

\bibitem{automatedremediation}
C.~Fernandez-Navarro and F.~Chicano, ``Automated llm-based accessibility
  remediation: From conventional websites to angular single-page
  applications,'' \emph{arXiv preprint arXiv:2602.17887}, 2026.

\bibitem{tabulara11y}
M.~Andruccioli, B.~Bassi, G.~Delnevo, and P.~Salomoni, ``The tabular
  accessibility dataset: A benchmark for llm-based web accessibility
  auditing,'' \emph{Data}, vol.~10, no.~9, p. 149, 2025.

\bibitem{llmweba11yslr}
W.~Aljedaani and R.~H. Mollik, ``Large language models for web accessibility: A
  systematic literature review,'' in \emph{Proceedings of the 23rd
  International Web for All Conference}, 2026, pp. 160--171.

\bibitem{aaa}
M.~Gu, Z.~Wang, S.~Lai, Z.~Gao, S.~Zhou, and J.~Bu, ``Towards scalable web
  accessibility audit with mllms as copilots,'' in \emph{Proceedings of the
  AAAI Conference on Artificial Intelligence}, 2026.

\end{thebibliography}

\end{document}